\documentclass[
twocolumn,
amsmath,
amssymb,
floatfix,
journal=prl
]
{revtex4-2}

\usepackage{graphicx}
\usepackage{dcolumn}
\usepackage{bm}
\usepackage[colorlinks,bookmarks=false,citecolor=blue,linkcolor=red,urlcolor=blue]{hyperref}

\usepackage[version=4]{mhchem}
\usepackage{comment}
\usepackage{siunitx}

\usepackage{natbib}
\usepackage{bibentry}
\usepackage[dvipsnames]{xcolor}
\usepackage{soul} 
\usepackage[caption=false]{subfig}
\usepackage{ulem}

\definecolor{orange}{rgb}{1.0, 0.3, 0}
\definecolor{drkgr}{rgb}{0.02,0.5,0.05}

\begin{document}
	
\title{
Resonating holes vs molecular spin-orbit coupled states in group-5 lacunar spinels
}

\author{Thorben Petersen\textsuperscript{1}}
\email{t.petersen@ifw-dresden.de}
\author{Pritam Bhattacharyya\textsuperscript{1}}
\author{Ulrich K. R{\"o}{\ss}ler\textsuperscript{1}}
\author{Liviu Hozoi\textsuperscript{1}}
\email{l.hozoi@ifw-dresden.de}

\affiliation{\textsuperscript{1}Institute for Theoretical Solid State Physics, 
Leibniz IFW Dresden, Helmholtzstra{\ss}e 20, D-01069, Dresden, Germany}

\date{\today}

\begin{abstract} 
The valence electronic structure of magnetic centers is one of the factors that determines the
characteristics of a magnet. 
This may refer to orbital degeneracy, as for $j_\text{eff}\!=\!1/2$ Kitaev magnets, 
or near-degeneracy, e.\,g., involving the third and fourth shells in cuprate superconductors. 
Here we explore the inner structure of magnetic moments in group-5 lacunar spinels, fascinating
materials featuring multisite magnetic units in the form of tetrahedral tetramers. 
Our quantum chemical analysis reveals a very colorful landscape, 
much richer than the single-electron, single-configuration description applied so far 
to all group-5 Ga$M_4X_8$ chalcogenides, 
and clarifies the basic multiorbital correlations on $M_4$ tetrahedral clusters:
while for V strong correlations yield a wave-function that can be well described 
in terms of four V$^{4+}$V$^{3+}$V$^{3+}$V$^{3+}$ resonant valence structures, 
for Nb and Ta a picture of dressed molecular-orbital $j_\text{eff}\!=\!3/2$ entities 
is more appropriate. 
These internal degrees of freedom likely shape vibronic couplings, phase transitions, 
and the magneto-electric properties in each of these systems. 

\end{abstract}


\maketitle

\section*{Introduction}
A magnet is a collection of magnetic moments; its characteristic properties are determined by the
nature of those moments and by how they mutually interact.
To shape the properties of magnetic materials according to specific requirements
we therefore need to
(i) understand and (ii) have some degree of control over magnetic moments --- inner structure and
the way they interact with each other.
It turns out that both --- inner morphology and mutual interaction --- depend on the subtle interplay
of electronic correlations, spin-orbit couplings (SOCs), and crystal-field effects (CFEs).
The combined action of these three factors received enormous attention in recent years.
New insights and new ideas in this research area have led to new physical models, new concepts, and
new research paths, as for example Kitaev's spin model \cite{Kitaev2006} and extensive associated work \cite{Takagi2019}.

\begin{figure}[b]
	\includegraphics[width=\columnwidth]{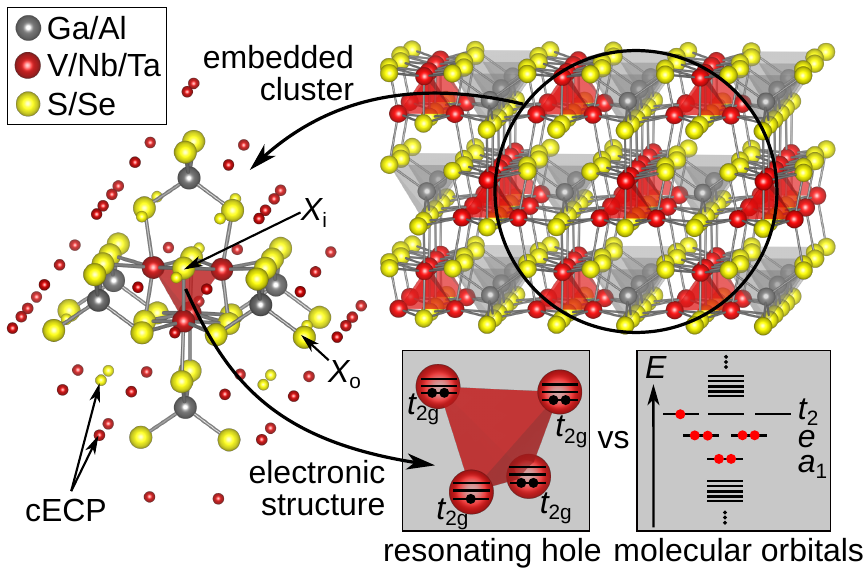}
	\caption{
		\textbf{Lacunar spinel crystal structure and embedded cluster model employed 
		in this study.}
		From the extended solid, a [$M_4X_{28}$Ga$_6$]$^{25-}$ cluster is cut 
		for quantum chemical analysis ($M$ = V, Nb, Ta and $X$ = S, Se).
		Each transition metal ion $M$ is surrounded by six ligand ions $X$ 
		in a distorted octahedron.
		$X_\text{i}$ and $X_\text{o}$ labels indicate $X$ atoms inside and outside the 
		$M_4$ tetrahedron, respectively, referring to the basis set assignment 
		in Supplementary Table\:2.
		The small spheres indicate embedding capped effective core potentials (cECPs).
		Inset: valence level representations analyzed in this study.
		}
\label{fig:cluster}
\end{figure}

Here we reveal what lies behind effective moments in each of the group-5 
Ga$M_4$$X_8$ lacunar spinels ($M$ = V, Nb, Ta and $X$ = S, Se),
fascinating materials displaying remarkable magnetic \cite{Kezsmarki2015}, magneto-electric 
\cite{Geirhos2021b,Geirhos2021}, and transport \cite{AbdElmeguid2004,Pocha2005} properties.
The characteristic structural feature of this family of compounds is that the transition-metal 
ions are clustered as $M_4$ tetrahedra (see Fig.\:\ref{fig:cluster}).
The latter can be then viewed as effective (magnetic) sites of a $fcc$ lattice 
and their electronic structure can be described in terms of $T_\text{d}$ point-group 
symmetry-adapted cluster orbitals ---
$a_{1}/a_{2}$, $e$ and $t_{1}/t_{2}$.
From electronic-structure calculations based on density functional theory (DFT), 
an $a_1^2e^4t_2^1$	valence electron configuration was inferred \cite{Pocha2000,Mueller2006,Kim2014}.
While indications of genuine many-body physics are available from both {\it ab initio} quantum chemical investigations \cite{Hozoi2020,Petersen2022,Petersen2023} and dynamical mean field theory 
(DMFT) \cite{Kim2020}, an in-depth profile of correlation effects across the 
3$d$-4$d$-5$d$ lacunar-spinel series is missing, which is the scope of our present 
quantum chemical study.

Besides clarifying essential electronic-structure features, the multiconfigurational
wave-function analysis that we provide --- 
in terms of either localized, site-centered or multisite orbitals --- 
makes these materials a distinct correlated-electron model system, as illustrative 
but much more captivating than other platforms typically employed 
to illustrate electronic correlations, as e.\,g., the \ce{H2} molecule 
for variable interatomic separation \cite{RamosCordoba2016,Izsak2023}.
Using as indicator for the strength of correlations the weight of ionic configurations in the
ground-state wave-function, we picture 
(i) what strong correlations mean in the 3$d$ vanadates (GaV$_4$S$_8$, GaV$_4$Se$_8$, AlV$_4$S$_8$) and
(ii) the notion of moderate correlations and `dressed'
$j_{\mathrm{eff}}\!=\!3/2$ objects in the 4$d$ (GaNb$_4$S$_8$, GaNb$_4$Se$_8$) and 5$d$ (GaTa$_4$Se$_8$)
variants. 
Even for the heavier cations, when expressing the multiconfigurational wave-functions 
in terms of delocalized multisite orbitals, the weight of the $(...)t_2^1$ configuration
presently assumed to correctly describe the ground state amounts to only 60\%.
Impressively, that shrinks to as little as 20\% for 3$d$ electrons.
Yet, SOCs are still effective --- even in the vanadates, those give rise to a
spin-orbit-induced splitting of $\approx$10\:meV for the ground-state term.
Also peculiar here is the near-degeneracy of the ground and a higher-spin state,
which should be possible to evidence by either spectroscopy or pressure experiments.
All these electronic-structure features for the 3$d$ case --- massive correlations, scaled down but
still detectable spin-orbit fine structure, and close proximity of high-spin states --- outline a few
important differences between 3$d$ and 4$d$/5$d$ group-5 lacunar spinels, i.e., the starting point in
understanding the differences in their magnetic properties.

\section*{Results}
\subsection*{High-temperature, tetrahedral-symmetry multiplet structure.\,}

\begin{figure*}[htbp]
  \includegraphics[width=0.8\textwidth]{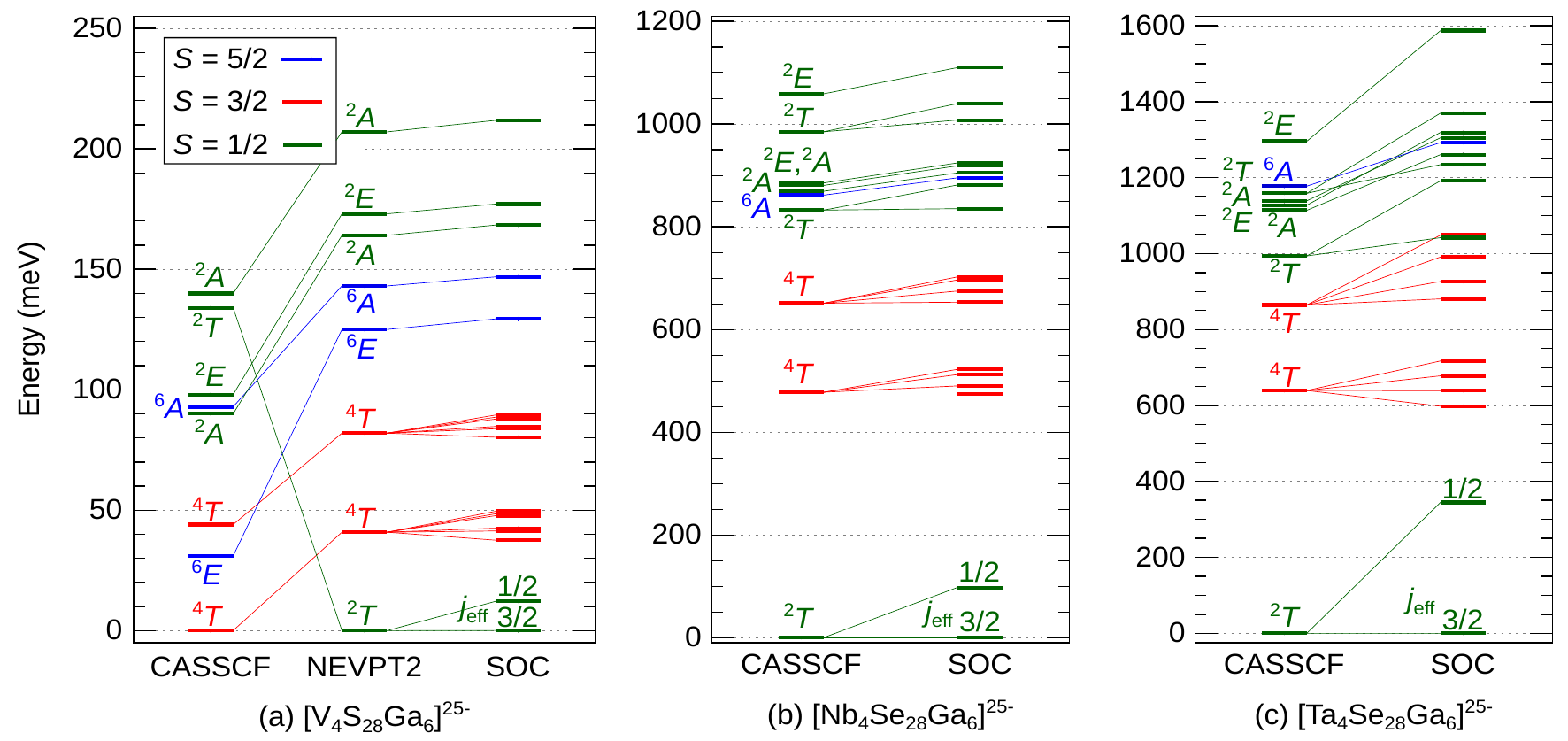}
  \caption{
	\textbf{
	Low-energy multiplet structures
	for \ce{GaV4S8}, \ce{GaNb4Se8}, and \ce{GaTa4Se8}.}
    The excitation energies were calculated for (a) \ce{[V4S28Ga6]^25-}, 
    (b) \ce{[Nb4Se28Ga6]^25-}, and (c) \ce{[Ta4Se28Ga6]^25-} embedded clusters (CAS(7e,12o)).
	Different spin states $S$ ($S$ = 1/2, 3/2, 5/2) are shown in green, red, and blue, respectively.
	SOC was accounted for on top of the CASSCF/NEVPT2 wave-functions.
    The corrections brought by NEVPT2 are minor for the Nb- and Ta-based compounds and 
    therefore not depicted (see Supplementary Tables 4 and 5).
  }
  \label{fig:m4x16ga6x12_f43m}
\end{figure*}

The $M_4$-tetrahedron multiplet structure, as computed for the high-temperature $F\bar{4}3m$ cubic
lattice arrangement of group-5 lacunar spinels, is displayed in Fig.\:\ref{fig:m4x16ga6x12_f43m}.
The multiconfigurational complete active space self-consistent field (CASSCF) method \cite{Roos1987}
with an active space of seven electrons in twelve orbitals [(7e,12o)-CAS] was applied.
Those orbitals are depicted in Supplementary Figure\:1.

For \ce{GaV4S8} (Fig.\:\ref{fig:m4x16ga6x12_f43m}.(a)\,), this (7e,12o)-CASSCF yields 
a high-spin ($S\!=\!3/2$) ground state.
Accounting for dynamical correlation effects in the scheme of \textit{post}-CASSCF $N$-electron
valence perturbation theory (NEVPT2) \cite{Angeli2001} corrects this state ordering.
Near degeneracy of low- and high-spin states is an effect often seen in 3$d$ systems, due to the 
similar magnitude of Coulomb interactions and various valence level splittings; 
in solid-state context, a well-known example is LaCoO$_3$ 
(see \cite{Hozoi2009} for a quantum chemical investigation).
Spin-crossover effects were also observed in DMFT calculations on \ce{GaV4S8} \cite{Kim2020}.
In contrast, for the Nb- (Fig.\:\ref{fig:m4x16ga6x12_f43m}(b)) and Ta-based materials
(Fig.\:\ref{fig:m4x16ga6x12_f43m}(c)), the CASSCF(7e,12o) methodology already ensures
a reasonably good description ---
the NEVPT2 scheme provides only minor corrections to the relative energies 
in the 4$d$ and 5$d$ systems (see Supplementary Tables\:4 and 5).

While a $^2T_2$ state is found as ground state in all compounds, the $a_1^2e^4t_2^1$ electronic
configuration (where $a_1$, $e$, and $t_1$ are symmetry-adapted, molecular-like orbitals in
$T_\text{d}$ point-group symmetry) contributes with weights well below 100\% to the ground-state
wave-functions; this aspect is discussed in detail in the following section.
SOC further splits the degenerate $^2T_2$ components into a $j_\text{eff}=3/2$ ground 
and a $j_\text{eff}=1/2$ excited state in all instances, 
with the magnitude of this splitting increasing from 12\:meV ($3d$) to 97\:meV (4$d$) 
and 345\:meV ($5d$ ions);
the latter number, in particular, suggests that the origin of the peak found at $\approx$0.3\:eV
in resonant inelastic X-ray scattering (RIXS) measurements on \ce{GaTa4Se8} \cite{Jeong2017}
is spin-orbit splitting within the $^2T_2$ ($a_1^2e^4t_2^1$) levels, different from the initial
interpretation in terms of $e^4t_2^1$ $\rightarrow$ $e^3t_2^2$ transitions
\cite{Jeong2017,Petersen2022}.

It is seen that the separation between the ground $^2T$ and the first excited $^4T$ state 
also increases for heavier transition-metal species:
41\:meV in \ce{GaV4S8}, 475\:meV in \ce{GaNb4Se8}, and 598\:meV in \ce{GaTa4Se8}.
A denser set of low-lying excited state levels in the vanadate could explain the significant
deviation from Curie-Weiss behavior seen at higher temperatures in susceptibility measurements
\cite{Widmann2017,Stefancic2020}, although a quantitative analysis would require the inclusion 
of vibronic couplings and intersite magnetic interactions (see discussion in 
Section V of the Supplementary Information and in our previous work \cite{Petersen2022}).
The latter then dictate the structure of the inelastic neutron scattering spectra 
\cite{Pokharel2021}.


\subsection*{Ground-state correlations in cubic group-5 spinels}

A major difference between how the single-tetrahedron electronic structure is presently depicted in 
the literature and in our quantum chemical results is the composition of the ground-state $^2T_2$ term.
Different from the 100\% $a_1^2e^4t_2^1$ ground state assumed so far on the basis of 
DFT computations for these materials, we find weights of 72\% in \ce{GaTa4Se8}, 
65\% in \ce{GaNb4Se8}, and as little as 21\% in \ce{GaV4S8} for the $a_1^2e^4t_2^1$ configuration.
Other configurations contributing to the ground-state wave-functions imply, for example, double
excitations into higher-lying $t_1$ and $t_2$ levels, each of those configurations with a weight
of a few percents or even less (see also Supplementary Table\:6).

The much more pronounced multiconfigurational character of the vanadate ground-state
wave-function in the symmetry-adapted orbital basis is also reflected in the correlation 
index proposed by Ramos-Cordoba \textit{et al.}~\cite{RamosCordoba2016}, 
which serves as a measure for the extent of near-degeneracy effects 
(also referred to as nondynamical correlation in quantum chemistry). 
Using the (7e,12o) CASSCF natural-orbital occupation numbers, we derived nondynamical
correlation indices $I_\text{ND}$ \cite{RamosCordoba2016} ranging 
from $\approx$2 in vanadates (1.96 for \ce{GaV4S8} and 2.00 for \ce{GaV4Se8}) to 
$\approx$1.05 in the 4$d$ variants (1.05 for \ce{GaNb4S8} and 1.07 for \ce{GaNb4Se8}) 
and 0.92 in \ce{GaTa4Se8} (details are given in the Supplementary Table\:7).
To put this in perspective, along the H$_2$ dissociation curve, $I_\text{ND}$ evolves 
from less than 0.1 at equilibrium distance to 0.5 towards dissociation \cite{RamosCordoba2016},
although a direct comparison of $I_\text{ND}$ values between chemically different systems
is not straightforward.

\begin{figure}[htbp]
	\includegraphics[width=0.7\columnwidth]{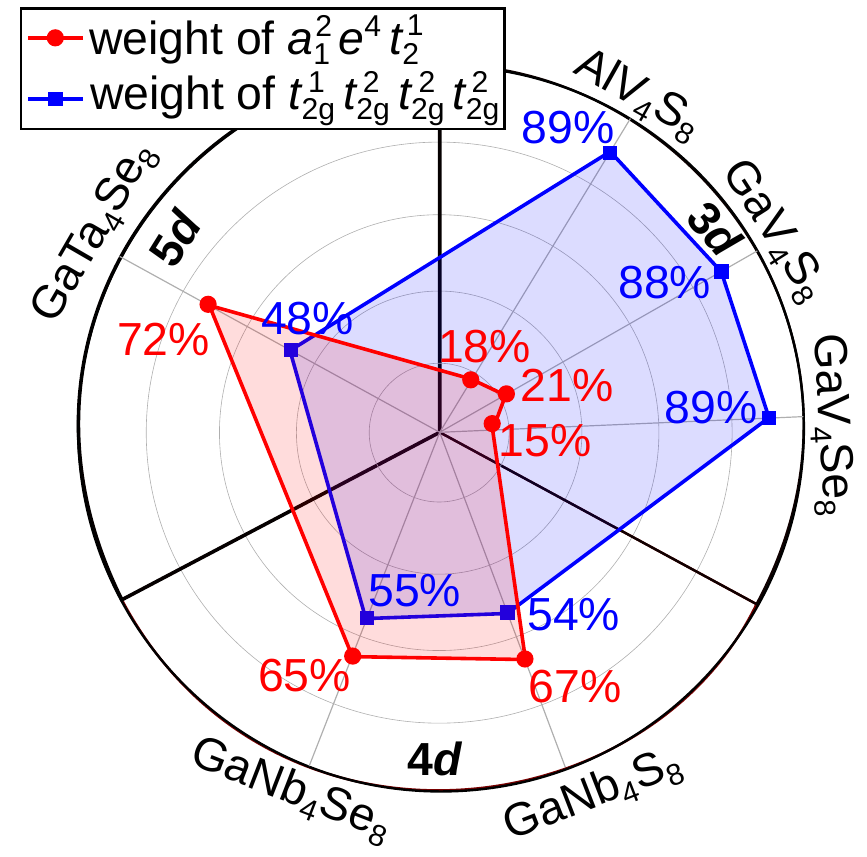}
	\caption{
		\textbf{
		Polar plot with weights for the leading configuration in symmetry-adapted 
		(red) and localized (blue) orbital basis.
		}
		For each system, the analysis is performed in terms of ground-state optimized 
		CASSCF(7e,12o) wave-functions.
	}
	\label{fig:GS-config_f-43m}
\end{figure}

Looking for further insight, we re-expressed the many-body ground-state wave-functions 
in terms of atomic-like, single-site orbitals.
The orbital localization module available in \textsc{Orca} was employed for this purpose; 
from the 12 symmetry-adapted orbitals (1$\times$$a_1$, 1$\times$$e$, 1$\times$$t_1$,
2$\times$$t_2$) obtained in the (7e,12o) CASSCF we arrive then to 
12 $t_\text{2g}$-like functions (three per pseudo-octahedrally coordinated transition-metal ion, 
see Supplementary Figure\:1).
For both orbital bases, symmetry-adapted or site-centered, the weights of the 
leading configurations are illustrated in Fig.\:\ref{fig:GS-config_f-43m},
for \ce{GaV4S8}, \ce{GaV4Se8}, \ce{GaNb4S8}, \ce{GaNb4Se8}, \ce{GaTa4Se8}, 
and an $A$-site-substituted member of the family (see also Supplementary Table\:6).
Interestingly, a low weight of the $a_1^2e^4t_2^1$ configuration in the symmetry-adapted orbital 
basis is associated with large weight of the $t_\text{2g}^1t_\text{2g}^2t_\text{2g}^2t_\text{2g}^2$
(i.\,e., V$^{4+}$V$^{3+}$V$^{3+}$V$^{3+}$) configurations in the localized-orbital representation; 
for the \ce{AlV4S8}, \ce{GaV4S8}, and \ce{GaV4Se8} vanadates, in particular, 15-20\% $a_1^2e^4t_2^1$
translates into 85-90\% configurations of $t_\text{2g}^1t_\text{2g}^2t_\text{2g}^2t_\text{2g}^2$ type.
The remaining part stems mainly from triply-occupied transition-metal centers, 
i.\,e., excited-state configurations of $t_\text{2g}^1t_\text{2g}^2t_\text{2g}^1t_\text{2g}^3$
(V$^{4+}$V$^{3+}$V$^{4+}$V$^{2+}$) type.
Weights of only $\approx$9\% for the latter 
indicate much stronger correlations in the case of V-based lacunar spinels:
intersite fluctuations are heavily suppressed, compared to the 4$d$ and 5$d$ compounds.
In a Mott-Hubbard picture, stronger localization is the result of larger $U/t$ ratio.
In other words, correlations are moderate in the 4$d$ and 5$d$ compounds and strong in the vanadates
--- for the latter, an expansion in terms of four V$^{4+}$V$^{3+}$V$^{3+}$V$^{3+}$ resonant
valence structures already provides a reasonably good description.
For perspectives on valence bond theory, the reader is referred to 
e.\,g. \cite{VB_shaik_03,VB_truhlar_07}.

\section*{Discussion}
To analyze in detail how correlations evolve from 3$d$ to 4$d$ and 5$d$ ions for the same type of
leading ground-state configuration and in the same crystallographic setting is difficult.
3$d^5$ (Mn$^{2+}$, Fe$^{3+}$) species, for example, tend to adopt a $t_\text{2g}^3e_g^2$ ground-state
electron configuration, while 4$d^5$ (Ru$^{3+}$, Rh$^{4+}$) and 5$d^5$ (Ir$^{4+}$) varieties 
display a $t_\text{2g}^5$ valence-orbital occupation.
Thinking of lower $d$-shell filling, Mo$^{5+}$ 4$d^1$ and Os$^{7+}$ 5$d^1$ ions, for instance, 
can be found in double-perovskite $fcc$ settings \cite{Chen2010}, but that is not the case 
for Ti$^{3+}$ or V$^{4+}$ 3$d^1$.

Here we individualize the group-5 lacunar spinels as a unique platform that makes it possible to
illustrate how correlations shape many-body wave-functions across a given group of the $d$ block
--- 3$d$ to 4$d$ and 5$d$, for the same kind of leading electron configuration and in the same
crystallographic setting.
In particular, we provide new, important insights by expressing the many-body $M_4$-tetrahedron
wave-function in terms of localized single-ion $t_\text{2g}$ orbitals.
We show that in the vanadates strong correlations yield a weight of 85--90\% for the
$t_\text{2g}^1t_\text{2g}^2t_\text{2g}^2t_\text{2g}^2$ (i.\,e., V$^{4+}$V$^{3+}$V$^{3+}$V$^{3+}$) configurations;
ferromagnetic double exchange shows up in this setting and yields near degeneracy of the low-lying
low- and high-spin states --- the $S\!=\!1/2$ doublet is obtained as single-tetrahedron ground-state
term only through a more advanced many-body treatment.
In contrast, smaller $U/t$ ratios in the 4$d$ (Nb) and 5$d$ (Ta) systems take us away from the regime
of strongly correlated electrons: in localized-orbital basis, we see that stronger charge fluctuations
reduce the weight of $t_\text{2g}^1t_\text{2g}^2t_\text{2g}^2t_\text{2g}^2$ resonances to $\sim$50\%;
in molecular-orbital representation, significantly larger hoppings ($t$) and consequently larger
bonding-antibonding splittings make that the {\it Aufbau} principle is to first approximation usable,
with weights in the range of 65-75\% for the $a_1^2e^4t_2^1$ configuration.
 
This result suggests a physical picture for the nonmagnetic states found in the Nb- and 
Ta-based lacunar spinels: as the $M_4$-cluster electrons are prone to stronger fluctuations with
larger spatial spread,	inter-cluster couplings are able to create ``valence bond'' spin singlets, 
as concluded from experiment \cite{Waki2010,Ishikawa2020,Yang2022}.
The peculiar pseudospin structure must play a role in the superconductivity mechanism 
under pressure, which is speculated to be unconventional owing to closeness of magnetic 
states and spin fluctuations.

While it is clear that the on-site correlations affect the magnetic properties, they are only 
indirectly discernible in available experimental data, as already pointed our for susceptibility 
\cite{Widmann2017, Stefancic2020} and inelastic neutron scattering measurements \cite{Pokharel2021} 
above.
More direct experimental insight into the details of the single-tetrahedron correlated electronic
structure might be derived from resonant inelastic x-ray scattering experiments, as in the case 
of other clustered compounds, either $d$- \cite{Gu2022} or 
$p$-electron \cite{Denlinger2002a,Denlinger2002b,Petersen2022b} based.

Overall weights of $\gtrsim$50\% for the $t_\text{2g}^1t_\text{2g}^2t_\text{2g}^2t_\text{2g}^2$ 
resonant valence structures suggest the $t$-$U$-$V$ model (or $U$-$tt'$-$VV'$ variants, 
where the primes denote inter-tetrahedral
hopping matrix elements and Coulomb repulsion integrals) as means to explore correlation-induced 
symmetry breaking.
Such numerical investigations could provide qualitative insights into the different types of 
low-temperature lattice symmetries realized in lacunar spinels and also into the polar properties 
of these materials.
An analysis in terms of only three molecular-like $t_2$ orbitals and one electron as in e.g. 
Ref. \cite{Dally2020} does not seem promising: 
according to our data, especially in the vanadates, the ($a_1^2e^4$)$t_2^1$ description is not
appropriate --- the $M_4$ tetrahedron $t_2$ electron cannot be separated from the other six $d$-ion
valence electrons, symmetry breaking should be rather described in terms of resonating holes
in localized orbital basis (i.e. V$^{4+}$ `holes' in V$^{3+}$ `background').

In sum, our quantum chemical data provide unparalleled specifics as concerns the correlated
electronic structure of the group-5 lacunar spinels, well beyond the featureless $a_1^2e^4t_2^1$
picture circulated so far in the literature.
Stronger correlations in the vanadates imply substantially less weight for the $a_1^2e^4t_2^1$
configuration as compared to the Nb and Ta compounds
and render the molecular-orbital picture \cite{Pocha2000,Mueller2006,Browne2020} inappropriate.
Remarkably, spin-orbit coupling is still effective, even for the vanadates --- the predicted fine
structure with a splitting of $\approx$10 meV between the $j\!\approx\!3/2$ and $j\!\approx\!1/2$
terms should be detectable experimentally.
The stronger intersite fluctuations ($M^{3+}M^{3+}$$\rightarrow$$M^{2+}M^{4+}$) and the larger
weight (65-75\%) of the $a_1^2e^4t_2^1$ molecular-orbital
configuration in the Nb and Ta spinels indicate that the 4$d$ and 5$d$ systems 
are closer to the Hartree-Fock limit.
The different nature of valence-space charge fluctuations across the group-5 family 
of lacunar spinels should be relevant to inter-tetramer exchange; on-site charge fluctuations, 
for example, were shown to strongly renormalize intersite exchange 
in cuprate superconductors \cite{Bogdanov2022}.
Assessing cooperative effects in group-5 lacunar spinels through calculation of inter-cluster
couplings will require approaches able to incorporate the correlated nature of the $M_4$-cluster
ground states.


\section*{Methods}

The basic building block in the lacunar-spinel structure was described by
a \ce{[$M$4$X$28Ga6]^25-} embedded cluster model ($M$ = V, Nb, Ta; $X$ = S, Se) 
(see Fig.\:\ref{fig:cluster}).
Experimentally determined high-temperature lattice parameters were adopted, as reported by Stefancic
\textit{et al.}~\cite{Stefancic2020} for \ce{GaV4S8} and by Pocha \textit{et al.}~for \ce{GaNb4Se8}
and \ce{GaTa4Se8} \cite{Pocha2005}.
The influence of the surrounding bulk atoms was modeled by a finite 
point charge field (PCF) generated through the \textsc{Ewald} program 
\cite{Klintenberg2000,Derenzo2000}.
A buffer region of $60$ capped effective core potentials (cECPs) was set up between the 
quantum cluster and PCF (indicated by the smaller atoms in Fig.\:\ref{fig:cluster})
(for details, see Section I of the Supplementary Information).

As initial step in our study, quasi-restricted orbitals (QROs \cite{Neese2006}) were generated 
from an unrestricted Kohn-Sham B3LYP calculation for a single-configuration $S$\:=\:5/2 state 
with initial-guess orbitals from Hueckel theory.
The Hueckel guess ensures that the QROs fulfill $T_\text{d}$ point-group symmetry.
Subsequently, 12 [$M_4$]$^{13+}$ molecular orbitals around the HOMO-LUMO gap
were identified from the QROs and used as a starting point for complete active space
self-consistent field (CASSCF) \cite{Roos1987} calculations.
Major convergence problems as encountered in earlier quantum chemical studies \cite{Hozoi2020} are
circumvented in this way.
The valence-space multiplet structure was derived from state averaged (SA) CASSCF optimizations 
with those 12 orbitals and seven valence electrons defining the active space 
(denoted in quantum chemistry as (7e,12o) CASSCF), consequently corrected for dynamical
correlation by \textit{N}-electron valence 2\textsuperscript{nd} order perturbation theory (NEVPT2)
\cite{Angeli2001}.
Both methods were accelerated by the resolution of identity (RI \cite{Neese2003}) and chain-of-spheres
(COS \cite{Neese2009}) approximations for Coulomb and exchange integrals 
with automatically generated auxiliary basis sets \cite{Stoychev2017}.
All calculations were done using the program package \textsc{Orca}, v5.0.3 \cite{Neese2022}.

\subsection*{Data Availability}
The quantum chemical data (coordinates of quantum clusters and point charge fields, \textsc{Orca} outputs, and magnetic susceptibility simulations) generated in this study 
have been deposited in the RADAR database under the DOI \href{https://doi.org/10.22000/1655}{10.22000/1655}.

%

\vspace{2em}
\subsection*{Acknowledgments}
We thank I. Kézsmárki, P.~Fulde and R.~C.~Morrow for discussions 
and U.~Nitzsche for technical assistance.
This work was supported by the German Research Foundation (Deutsche Forschungsgemeinschaft, DFG),
Project No.~437124857.
P.B. acknowledges funding from the DFG, Project No.~441216021.

\vspace{2em}
\subsection*{Author Contributions}
T.P. carried out the quantum chemistry calculations with assistance from 
P. B., U.K.R. and L.H.
All authors were involved in writing the manuscript.
U.K.R. and L.H. planned the project.

\vspace{2em}
\subsection*{Competing Interests}
The authors declare no competing interests. 

\subsection*{Supplementary Information}
The online version contains supplementary material available at
\href{https://doi.org/10.1038/s41467-023-40811-y}{https://doi.org/10.1038/s41467-023-40811-y}.


\end{document}